\documentclass[usenatbib]{aa}  
\usepackage{graphicx}
\usepackage{natbib}
\usepackage{amssymb}
\usepackage{amsmath}
\def\pl{{\sc pl}}

\def\bb{{\sc bb}}

\def\bb{{\sc bb}} 
\def\I{{\em INTEGRAL}} 
\def\BS{{\em Beppo}SAX} 
\def\R{{\em RXTE }} 
\def\4u{4U~1954+319} 
\def\be{\begin{equation}} 
\def\ee{\end{equation}} 
 
\begin{document} 
   \title{A new symbiotic low mass X-ray binary system: 4U~1954+319} 
 
 \author{F. Mattana\inst{1,2}\fnmsep\thanks{\email{mattana@iasf-milano.inaf.it}}, 
 D. G\"otz\inst{3}, M. Falanga\inst{3}, F. Senziani\inst{1,4,5}, A. De
 Luca\inst{1}, P. Esposito\inst{1,4}, \and P.A. Caraveo\inst{1}
} 
 
\offprints{F. Mattana} 
\titlerunning{4U~1954+319}  
\authorrunning{F. Mattana, D. G\"otz, M. Falanga, et al.}  
  
\institute{INAF --Istituto di Astrofisica Spaziale e Fisica Cosmica, via 
Bassini 15, 20133 Milano, Italy 
\and Universit\`a di Milano Bicocca, Dipartimento di Fisica G. Occhialini, 
P.za della Scienza 3, 20126 Milano, Italy 
\and CEA Saclay, DSM/DAPNIA/Service d'Astrophysique (CNRS FRE  
  2591), F-91191, Gif sur Yvette, France 
\and Universit\`a di Pavia, Dipartimento di Fisica
Nucleare e Teorica and INFN-Pavia, via Bassi 6, I-27100 Pavia, Italy    
\and Universit\'e Paul Sabatier, 31062 Toulouse, France
             } 
 
   \date{ } 
 

\abstract
	{}
{4U 1954+319 was discovered 25 years ago, but only recently has a clear picture of its nature 
begun to emerge. 
We present for the first time a broad-band spectrum of the source and 
a detailed timing study using more than one year of monitoring data.}
{The timing and spectral analysis was done using
publicly available {\em Swift}, \I, \BS, and \R/ASM data in the 
0.7 to 150 keV energy band.}
{The source spectrum is described well  by a highly absorbed (N$_{H}\sim$10$^{23}$ cm$^{-2}$) power law with a
high-energy exponential cutoff around 15 keV. An additional black body
component is needed below 3 keV to account for a soft excess. The derived
$\sim$5 hr periodicity, with a spin-up timescale of $\sim$ 25 years, could be
identified as the neutron star spin period.
The spectral and timing
characteristics indicate that we are dealing both with the slowest established 
wind-accreting X-ray pulsar and with the second confirmed member of the emerging
class dubbed "symbiotic low mass X-ray binaries" to host a neutron star.}
  {}

\keywords{binaries: close -- pulsars: individual (4U~1954+319) -- 
               stars: neutron -- X-ray: binaries} 
 
   \maketitle

\section{Introduction} 
\label{sec:intro} 
 
Since its discovery by the {\em ARIEL} satellite in 1981, 
\4u\ has appeared as a flaring X-ray source
\citep{w81}. Located in the Cygnus region, this source was 
observed by different X-ray telescopes
\citep[][]{forman78,cook84,w88,tweedy89,voges99}. 
From the observed spectral behaviour, the source was tentatively classified as a high-mass X-ray 
binary system hosting a neutron star (NS). However, little was known about 
the system, as the optical counterpart was not identified due to the 
ambiguous determination of the source position. The companion star 
was then classified as a close M-type giant or a distant and 
reddened Be star \citep[][]{cook84,tweedy89}. 
 
Recently, using the {\em Chandra} refined source position, 
\citet{masetti06} used optical spectroscopy to identify its companion
as an M4-5 III star, which is located within 1.7 kpc. 
Such identification suggests the system is composed of 
a compact object accreting through the wind of its M-type giant 
companion star.
This makes \4u\ the third low mass X-ray binary (LMXB) hosting an NS
and a late type giant companion, after GX~1+4  \citep[e.g.][]{chro97} and possibly 4U 1700+24
\citep{gaudenzi99}, for which however no coherent pulsation has been
reported to date \citep[e.g.][]{masetti02}. 
Therefore \4u\ could be attributed to the so-called ``symbiotic 
X-ray binaries'', an emerging subclass of LMXBs.
These systems with an evolved giant donor are the 
probable progenitors of most wide-orbit LMXBs \citep{chro97}.

Only recently has a periodic signal been reported from this source,  
whit a period of $\sim 5.09$ hours quasi-monotonically decreasing \citep{corbet06}.
If this period can be attributed to the 
spin, \4u\ would be the slowest binary NS known, with the possible exception of the enigmatic source
1E 161348-5055 in the Supernova remnant RCW 103 \citep{deluca06}. 
In this Letter we used several high-energy observations to unveil 
the nature of \4u\ through spectroscopic and timing analysis.

\section{Observations and data analysis} 

The {\em INTEGRAL} data set was obtained 
using the 2003/2005 publicly available observations 
within $<12^{\circ}$ from the source direction. 
We analysed 613 pointings from the IBIS/ISGRI coded mask telescope
\citep{u03,lebr03} at energies between 18 and 200 keV, for a total exposure 
time of 1.4 Ms. 
The data reduction was performed using the 
standard Offline Science Analysis (OSA) version 5.1. 
We extracted the 20--60 keV band light curve based on 
the single pointings.
During the \I~ observation, \4u\ was mainly in a quiescent phase (52717--53144 MJD), and in  
the last part of the observation, it entered a strong outburst phase 
(53312--53708 MJD) with an increase in count rate of a factor 40.
 During the outburst phase, \4u\ is clearly detected 
in the mosaic image with a significance level of $\sim45\sigma$. 
During the quiescent phase, \4u\ 
was not detected at a statistically significant level in the ISGRI 
data. Therefore we excluded these data from the timing and spectral analyses.

We also included  for the spectral analysis the
publicly available data from the Narrow Field Instruments
(NFIs) on board \BS. 
The source \4u\ was observed on May 4, 1998 for a net exposure of 19 ks in
the LECS \cite[][]{lecs} and for 46 ks in the MECS \cite[][]{mecs}. 
The average spectrum of the source was extracted from a region 
of 6$^{\prime}$ in both instruments. The spectra were then 
rebinned in order to have at least 30 counts per channel and three 
channels per resolution element. Note that \BS~ observations were 
too short to enable the timing analysis.  

We also analysed  the {\em Swift}/BAT  \citep{swift} survey data, 
collected between December 2004 and April 2006. The analysis was performed 
using the HEASoft software (release v.6.0.5). For each single pointing 
(with typical integration time of 5 minutes), we extracted an image 
in the 15--50 keV energy range and evaluated the source count rate at the {\em Chandra}
position. Thus, we obtained a barycentered light curve containing $\sim23000$ data points.

\begin{figure}[htb]
\centering 
\includegraphics[width=8.5cm]{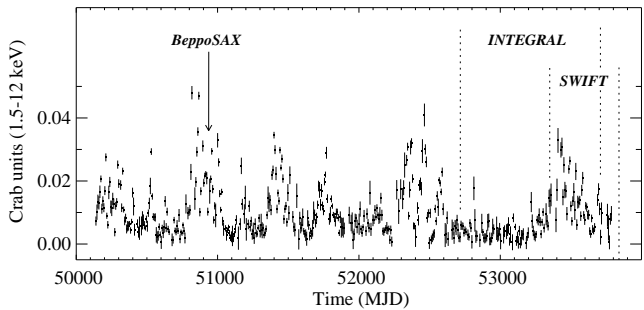} 
\caption{The \4u\ {\it RXTE}/ASM light curve in the 1.5--12 keV energy band 
(data-averaged over a 7-day interval).  We converted the ASM count rate to flux 
using 1 Crab $\approx 75$ cts/s \citep{l96}. } 
\label{fig:asm} 
\end{figure}

In Fig. \ref{fig:asm} we show the {\it Rossi X-ray Timing Explorer} ({\it RXTE}) {\it
All-Sky Monitor} (ASM) light curve of \4u, where we report 
the time interval in which the {\it INTEGRAL}, {\it Swift}, and \BS\ observations were performed.

\subsection{Spectral analysis} 
\label{sec:spectra} 
For the spectral analysis, we used ISGRI (18-150 keV), LECS (0.7--4.5 keV), 
and MECS (1.7--10 keV) data. The non-imaging \BS\ spectrometer PDS data
(15--300 keV)
could not be used due to the contamination of Cyg X--1. 
The analysis was done using XSPEC version 11.3 
\citep{arnaud96}. All spectral uncertainties in the results 
are given at a 90\% confidence level for single parameters.  

We first computed a  hardness-intensity diagram using the {\it RXTE/ASM}
data. The hardness is the ratio of the count rates in the 1.5--5
keV to the ones in the 5--12 keV energy band, and the intensity is the 1.5--12 keV count rate. Each
point
corresponds to $\sim 7$ days of
integration time. The source shows a significant linear
correlation (slope $\approx 1$) between flux and hardness.  
Therefore we searched for spectral changes in the ISGRI data, dividing
the observation into four different intensity levels. We fitted each
subset with a simple power-law (\pl) model, but it turned out to be
statistically inadequate ($\chi^{2}$/d.o.f = 43/12).   
The best fit was found by replacing the \pl\ with a cutoff \pl\ model thereby
obtaining a $\chi^{2}$/d.o.f = 10/11. We found the spectral parameters  
for the four data sets to be the same within the error bars, most
likely due to the low statistics during the flare,
so we did not observe the same hardness-intensity correlation as observed 
below 12 keV with the ASM data. 
The best-fit values for the
average ISGRI spectrum are found for a \pl\ photon index, $\Gamma$, of
$0.7\pm0.1$  and a high-energy cutoff $E_c \sim 14_{-3}^{+9}$ keV. 

We then fitted the \BS\ spectrum using a simple
photoelectrically-absorbed \pl\ model plus a black body
(\bb) model for the soft excess below 3 keV, as often observed in accreting pulsars
\citep[e.g.][]{hickox04} and a Gaussian emission line, resulting in a $\chi^{2}{\rm /d.o.f.} = 132/71$.
The best fit was found for a heavily absorbed spectrum with $N_{\rm
  H}=2.8\pm0.2\times 10^{23}$ cm$^{-2}$, a \bb\ temperature of
$kT$=0.11$\pm$0.01 keV to model the soft X-ray excess, a \pl\
index of $\Gamma=1.5\pm0.5$, $E_c \sim 17\pm12$ keV, and an iron emission line at $\sim$6.33$\pm$0.15 keV (with
the line width fixed at 0.5 keV). The quite high reduced $\chi^{2}$ indicates that the soft excess is
not perfectly modeled by the \bb\ component. The hydrogen
column density, $N_{\rm H}$, is found to be 30 times higher than the
Galactic value reported in the radio maps of \citet{dickey90}.

\begin{figure} 
\centering 
\includegraphics[angle=-90,width=8cm]{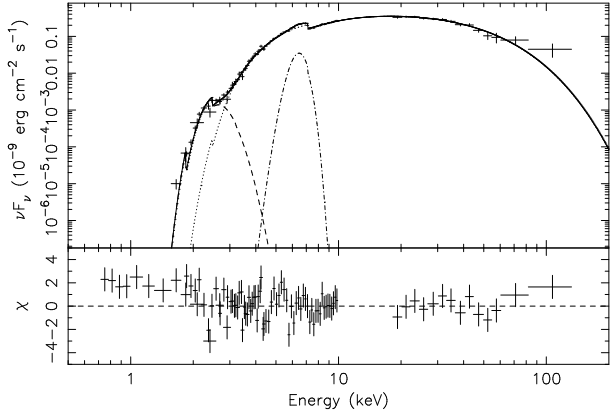} 
\caption{Broad-band spectrum of \4u, fitted with an absorbed \bb\,
  Gaussian line, and cutoff \pl. The data points correspond to the
  LECS/MECS (0.7--10 keV) and ISGRI (18--150 keV) spectra. 
  The lower panel shows the residuals with respect to the best-fit model.} 
\label{fig:spectrum} 
\end{figure}

The cutoff value is not well-constrained using the NFIs or ISGRI alone. This is due
to the fact that it falls outside of the energy ranges of the individual instruments.
In order to better constrain the broad-band (0.6--200 keV) spectral characteristics 
we jointly fitted the NFI and the ISGRI data. A multiplicative factor 
for each instrument was included in the fit to account for the uncertainty in the
cross-calibration of the instruments, as well as variability across the non-simultaneous 
observations. 
The best-fit parameters are 
$\Gamma$=1.1$\pm$0.1, $E_{c}$=17$\pm$2 keV, $kT$=0.123$\pm$0.004 keV, and a similar value of
$N_{H}$, with a $\chi^{2}/$d.o.f. of 138/83. 
The broad-band spectrum, together with the best-fit model, are shown in
Fig. \ref{fig:spectrum}.

\subsection{Timing analysis}
\label{sec:timing} 
 
For the timing analysis we used the BAT (15--50 keV) and ISGRI (18--50 keV) light curves
after solar-system barycentric correction. 

\begin{figure} 
\centering 
\includegraphics[angle=0,width=5.5cm,angle=-90]{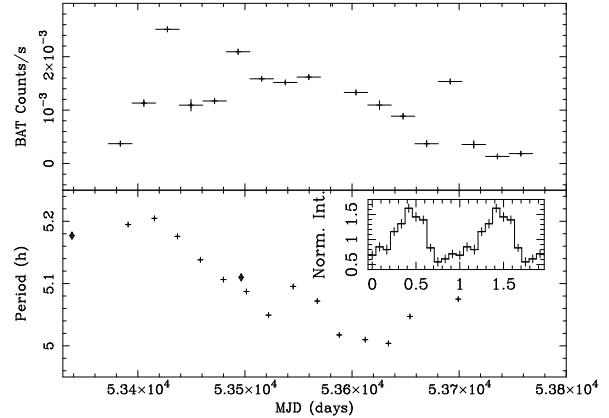} 
\caption{Upper panel: BAT light curve (15--50 keV) of \4u. The time bin size is 22 days. Lower panel: period evolution as measured with BAT (crosses) and ISGRI (diamonds). The inset represents the normalized BAT phase folded light curve between 53400 and 53500 MJD considering $P$ and $\dot{P}$. Two cycles are shown for clarity.} 
\label{fig:lc_period} 
\end{figure}

We searched for coherent pulsations of the source in the BAT data by
computing a power density spectrum (PDS) in the frequency  
range between    
2$\times$10$^{-8}$ and 10$^{-4}$ Hz from fast Fourier transforms. In the  
resulting PDS, a signal is evident at  $\nu = 5.45 \times 10^{-5}$ Hz. 
The peak is broad with an FWHM of 6.7$\times$10$^{-7}$ Hz, suggesting a period evolution.
Therefore the light curves were grouped into 100 cycles per time interval, and the best period 
was determined using an epoch-folding analysis. The distribution of the $\chi^2$ values versus
trial period were fitted as described in \citet{leahy87}.
The resulting 14 best-period values clearly show a spin-up trend, see
Fig. \ref{fig:lc_period}.

Using the ISGRI light curves, we confirmed the presence of a period at $\sim$5.17 hr.  
We attempted to search for a period change in the ISGRI data; but 
due to the sparse coverage using the 100 cycle time intervals, we could 
only derive 2 significant measurements (see Fig. \ref{fig:lc_period}), 
which are consistent with the previous findings.

Fitting the period evolution with a linear function, the derived spin-up 
trend is around $-$3$\times$10$^{-5}$ s s$^{-1}$, which
implies a spin-up time scale ($P/\dot P$) of $\sim$ 
25 years. However, we note that some scattering around this trend is present,
indicating a possible local spin-down. On the other hand, a much better fit 
was  obtained using a sinusoidal function, indicating a possible orbital period of 385$\pm$4 days and a projected semimajor axis, $a_x \sin i$, of 178$\pm$4 AU. However, since this  periodicity is of the order of the time interval spanned by the data, it is difficult to identify it with a possible orbital period (see below).

\section{Discussion} 
\label{sec:discussion} 
 
The timing and broad-band spectrum of \4u\ allowed us to investigate 
for the first time the nature of this system, which only recently has been tentatively 
attributed to the emerging NS ``symbiotic LMXBs'' class. 

The $\sim$5 hr measured periodicity could be due either to the system orbital
modulation or to the rotation of the compact object.
We can rule out that we are measuring the system orbital period.
In fact, given the typical luminosity and temperature of an M 4-5 III star \citep{lang92}, 
the stellar evolutionary models with solar  metallicities \citep{claret04} 
predict a mass of $\sim$1.2 M$_\odot$ for the companion.
Assuming a mass within 3 M$_{\odot}$ for the compact object, the orbital separation would be
much smaller than the donor star radius ($\sim$80 R$_\odot$). 
We are then left with the hypothesis of measuring the spin period of  
the NS hosted in the system \citep[a white dwarf would be excluded due to the excessive 
torque required, as already noted by][]{corbet06}. 

Furthermore, the observed spin-up could be attributed to the accretion
torque on the magnetized NS.
While suggestive of a sinusoidal trend (see Fig. \ref{fig:lc_period}
and Sect. \ref{sec:timing}), the time evolution of
the period cannot be accounted for by the Doppler shift of the
system's orbital motion: in this case the derived orbital parameters 
would imply a companion star mass of a few 10$^6$ M$_\odot$. Thus we can consider
$\sim$ 400 days a lower limit to the orbital period of the system. This implies a wide
orbit ($a >$2$\times$10$^{13}$ cm) and explains the fact that the optical 
spectrum reported by \cite{masetti06} shows no sign of the influence of the X-ray source.
  
We found a broad-band spectrum typical of accreting X-ray pulsars \citep[e.g.][]{joss84}
and, in particular, similar to the ones of GX 1+4 \citep{paul05}, and 4U 1700+24 \citep{tiengo05}.
The high column density found of the order of 3$\times$10$^{23}$ cm$^{-2}$ could be attributed to 
the local (circumstellar) absorption. In fact, this value is typical of the local environment of symbiotic
stars as indicated by observations of Rayleigh scattering in their UV spectra \citep{schmid97}.
The measured luminosity of $\sim 2 \times 10^{35}$ erg s$^{-1}$, assuming a 1.7 kpc source distance, 
points towards a wind-fed accretion system with the companion not filling its Roche lobe.
Taking the period, its derivative, and the luminosity into account, one
can use the standard accretion torque models \citep[][]{henrics} to derive an order-of-magnitude  
value for the dipolar magnetic field of the NS. It turns out
to be $B\sim 10^{12}$ G, which implies a magnetospheric  
radius $r_{m}=3\times 10^{8}\; \rm{cm}\; ( L_{X}/10^{37}\; \rm{erg\; s^{-1}})^{-2/7}(B/10^{12}\; \rm{G})^{4/7}\sim$ 10$^{9}$ cm, smaller than the corotation radius ($r_{co}=(GM_{X}P^{2}_{spin}/4\pi^2)^{1/3}\sim 1.2\times10^{11}$ cm, 
where $M_X$$\sim$$1.4$ $M_{\odot}$ is the mass of the NS), allowing the accretion process. We are hence dealing  
with a regular wind accreting X-ray  pulsar but with the longest known spin period.  
  
The very long period measured in \4u\ can be explained by taking into
account the age of the companion star \citep[$\sim$9$\times$10$^9$ yr,][]{claret04} and the standard 
binary pulsar evolutionary picture \citep{davies81}. In fact,
a newborn NS experiences a spin-down initially due to the magnetic dipole braking (radio phase).
In our case this phase lasts $\sim$8$\times$10$^9$ yr, namely the time the donor spends
on the main sequence. Since the orbit is quite wide, the NS can spin down practically as if
it was isolated. When the donor star leaves the main sequence, starting to evolve rapidly into
a red giant, the star mass loss increases and there is coupling between 
the infalling matter stopped by the magnetosphere and the magnetosphere itself  \citep[propeller phase,][]{illar75}.
By means of these processes during its whole lifetime, an NS with a magnetic field of a few 10$^{12}$ G can reach periods of the order of 10$^4$ s in $\sim$8.5$\times$10$^9$ yr. 

The distance where the ram pressure of the accretion flow balances the magnetic pressure
is called the Alfv\'en radius. A further increase in the accretion rate, as the donor becomes a red giant, 
allows the corotation radius to overcome the Alfv\'en radius and the NS to start accreting, becoming observable in X-rays and spinning-up. The emerging picture is an old, hence slow, NS 
accreting from the slow and dense wind of an evolved M type giant on a wide orbit. The inhomogeneities of the wind can explain the observed long-term X-ray
variability (see Fig. \ref{fig:asm}). The wide orbit also allowed the companion star not to be affected by the
Supernova explosion that generated the NS thereby enabling the donor star to follow its natural evolutionary track.

\section{Conclusions}
\label{sec:conclusions}

We analysed most of the recently available high-energy data of \4u\, 
deriving for the first time its broad-band (0.7--150 keV) spectrum
and analysing its timing characteristics. 
The luminosity and spin period derivative of the
system allowed us to infer a pulsar magnetic field of the
order of 10$^{12}$ G. In the framework of standard binary pulsar evolutionary models, 
this value, coupled to the age of the companion star, can 
justify the very long observed NS period without invoking any additional
spin-down process. 

It turns out that \4u\ is the
slowest accretion powered NS, and it is hosted in a ``symbiotic LMXB''
system. These are rare systems that could eventually evolve into 
wide-orbit binary systems hosting an NS and a white dwarf, but without mass transfer so hence 
very difficult to observe.

\begin{acknowledgements} 
FM acknowledges Fran\c{c}ois Lebrun and the CEA Saclay, DSM/DAPNIA/SAp, for hospitality during this work.
DG and MF acknowledge the French Space Agency  (CNES) for financial
support. PE is grateful to Albert Hoffmann for useful suggestions.
\end{acknowledgements}

\end{document}